\documentclass[preprint,12pt]{elsarticle}

\usepackage{graphicx,amsmath,bm}
\usepackage[dvips]{color}
\def\alf{Alfv\'en\,}
\def\bq{\begin{equation}}
\def\eq{\end{equation}}
\let\grad=\nabla
\def\cs{\cos\theta}
\def\drho{\delta\rho}

\def\persec{\mbox{s}^{-1}}
\def\b{\bf{b}}

\def\v{{{\bf{v}}}}
\def\va{v_A}

\def\vj{{{\bf{v}}}_j}

\def\vj{{{\bf{v}}}_j}

\def\vn{{{\bf{v}}}_n}
\def\B{{\bf{B}}}
\def\hB{\bf{b}}

\def\dB{{\delta\b}}
\def\J{{{\bf{J}}}}
\def\E{{{\bf{E}}}}
\def\k{{{\bf{k}}}}
\def\kh{\hat{{\bf{k}}}}

\def\dv{{\bf{\delta\v}}}

\newcommand\cross{{\bf{\times}}}
\def\curl{{\grad \cross}}
\newcommand{\delt} [1] {\frac{\partial #1}{\partial t}}

\newcommand{\tenq}[1]{\hbox{\oalign{$\bm{#1}$\crcr\hidewidth$\scriptscriptstyle\bm{\approx}$\hidewidth}}}

\journal{JATP}

\begin{document}

\begin{frontmatter}

\title{Dust modification of the plasma conductivity in the mesosphere}

\author[BP]{B.P.~Pandey\corref{cor1}}
\ead{birendra.pandey@mq.edu.au}
\author[SV]{S.V.~Vladimirov\fnref{fn1}}
\ead{sergey.vladimirov@sydney.edu.au}
\cortext[cor1]{Corresponding author}
\address[BP]{Department of Physics \& Astronomy, Macquarie University, Sydney 2109, NSW, Australia}
\address[SV]{Metamaterials Laboratory, National Research University of Information Technology, Mechanics \& Optics, St. Petersburg 199034, Russia and Joint Institute of High Temperatures, Russian Academy of Sciences, Izhorskaya 13/2, 125412 Moscow, Russia, School of Physics, The University of Sydney, Sydney 2006, NSW, Australia}





\begin{abstract}
Relative transverse drift (with respect to the ambient magnetic field) between the weakly magnetized electrons and the unmagnetized ions at the lower altitude ($< 80\,\mbox{km}$) and between the weakly magnetized ions and unmagnetized dust at the higher altitude ($< 90\,\mbox{km}$) gives rise to the finite Hall conductivity in the Earth{}\'s mesosphere. If, on the other hand, the number of free electrons is sparse in the mesosphere and most of the negative charge resides on the weakly magnetized, fine, nanometre sized dust powder and positive charge on the more massive, micron sized, unmagnetized dust, the sign of the Hall conductivity due to their relative transverse drift will be opposite to the previous case. Thus the sign of the Hall effect not only depends on the direction of the local magnetic field but also on the nature of the charge carrier in the partially ionized dusty medium.

As the Hall and the Ohm diffusion are comparable below $80\,\mbox{km}$, the low frequency ($\sim 10^{-4}—-10^{-5}\,\mbox{s}^{-1}$) long wavelength ($\sim 10^{3}—-10^{4}\,\mbox{km}$) waves will be damped at this altitude with the damping rate typically of the order of few minutes. Therefore, the ultra--low frequency magnetohydrodynamic waves can not originate below $80\,\mbox{km}$ in the mesosphere. However, above $80\,\mbox{km}$ since Hall effect dominates Ohm diffusion the mesosphere can host the ultra--low frequency waves which can propagate across the ionosphere with little or, no damping. 
\end{abstract}


\end{frontmatter}



\section{Introduction}
The Earth{}\'s ionosphere consists of horizontally stratified layers of partially ionized gas immersed in its magnetic field. The various altitude range of the ionosphere is divided for convenience as D, E and F layer, with the D region spanning between $60-90\,\mbox{km}$, the E region between $\sim\, 90-150\,\mbox{km}$ and the F region between $150-400\,\mbox{ km}$ \citep{K89}. The temperature from the ground up to $\sim 15\,\mbox{km}$ altitude (troposphere) decreases with height. The temperature rises in the stratosphere ($\in [15\,,50]\,\mbox{km}$) before decreasing again in the mesosphere where it has the lowest value at about $\sim 80-90\,\mbox{km}$. The temperature in the mesosphere is about $\sim 190\,\mbox{K}$ though $160\,\mbox{K}$ or even lower temperature is also possible at occasions. Surprisingly, the mesopause temperature in the polar regions is higher in winter than in summer.

The altitude profiles of the dominant neutrals as well as the ionized components and their variations above about $\sim 100\, \mbox{km}$ is quite well known.  However, the lower ionosphere ($\sim 60-100\,\mbox{km}$), owing to the limited experimental data base, to a large extent, is still poorly understood. The $60—-150\,\mbox{km}$ altitude region  is too high for balloons and too low for satellite observations posing considerable observational challenge, yet understanding of this region is crucial to the behaviour of radio transmission, the initiation of sprites above thunderstorm etc. Owing to the D-region\'s impact on the global climate change, this region has started receiving renewed attention \citep{B03}.
 
While the magnetic field shields Earth from the solar wind, meteoroids and dust freely penetrate the atmosphere. Meteors are observed at all altitude between $70-400\,\mbox{km}$ with small meteors evaporating between $70—-120\,\mbox{km}$ \citep{S13}. Meteoric smoke particles ($\sim \mbox{nm}$ in size; $1 \mbox{nm} = 10^{-9}\,\mbox{m}$) form from the recondensing of the ablated meteoroid material at $80—-90\,\mbox{km}$ \citep{RS61, H81}. The polar mesospheric summer echoes (PMSE) and noctilucent clouds (NLCs) also called polar mesospheric clouds (PMCs) are also observed at this altitude. There is a strong correlation between the observations of NLCs and PMSEs suggesting that they might have a common origin. The PMSE refers to the strong radar echoes observed at $50-1.3\,\mbox{GHz}$ due to electron scattering at Bragg scale whereas the NLC which is also observed in the similar range of frequencies refers to the formation of water ice particles \citep{CK93, CR97, RL04}. Typical density of these ice particles can vary between $\sim 10$ to $10^{3}\,\mbox{cm}^{-3}$ and their radius can vary between $\sim 10-—100\,\mbox{nm}$ \citep{RL04, FR09}. It is quite plausible that the size of the dust particles and the condensation of the water vapour in the NLCs are interlinked. For example, the presence of large dust can noticeably reduce the concentration of water vapour in the upper atmosphere and this in turn can decrease the particle sizes \citep{KVM05}.

The measurements during PMSE conditions shows a pronounced depletion of the electron number density.  In fact electron density can decrease by an order of magnitude where strong PMSE are observed. The depletion of electron density, usually called electron {\it bite--outs}, is a generic feature of the PMSE \citep{P69, U88}.  It would appear that the electron deficit at this altitude occurs only during PMSEs. However, the electron depletion is much more generic in the D-—region owing to the availability of meteoric smoke particles. In fact, there is a distinct deficit of electrons between $80$ and $90\,\mbox{km}$. This deficit can be explained by the presence of negatively charged meteoric smoke particles \citep{B13}.

Clearly, the plasma composition in the altitude region between $60-100\,\mbox{km}$ is chemically complex. Whereas toward the  
E—-region, the dominant ions are molecular ($\mbox{NO}^{+}$ and $\mbox{O}_2^+$), below a marked transition height cluster ions ($\mbox{H}^+\left(\mbox{H}_2\mbox{O}\right)_n$ and $\mbox{NO}^{+}\left(\mbox{H}_2\mbox{O}\right)_n$) constitute the bulk of the population \citep{K85, FT88}. The presence of charged and neutral dust (mesospheric smoke particles, ice particles) are also important constituent of the partially ionized gas. After forming at higher altitude ($\sim 90\,\mbox{km}$), the largest of dust particles settles to the lower altitude where they are visible as NLC whereas smaller dust is {\it visible} through strong radar echoes \citep{RL04, FR09, H96a, H96b}.  Estimates for the number density of small dust particles in the ionosphere ($\sim$ a few tens of nm to sub-visual in size) appears generally to be larger than about $10^2\,\mbox{cm}^{-3}$ under PMSE conditions \citep{H01}.

As noted above the dust in the D-region is either neutral or carry electronic charge. The ratio of charged to neutral dust particles is about 5 to 10 percent of the total dust number density \citep{G05, L05}   implying that the small dust particles are predominantly neutral. However, large ($> 20\,\mbox{nm}$) ice crystals are often negatively charged in the NLC. The presence of large dust can significantly augment the electron recombination rate causing the electron biteouts. The average charge on the dust is negative owing to the large mobility of electrons. The typical charge $Z$ on the dust will be $-1\,e$ for particles with radius $a \in [1, 10]\,\mbox{nm}$, $-2\,e$ for $30\,\mbox{nm}$ particles and $-3\,e$ for $50\,\mbox{nm}$ particles \citep{FR09}. Here $e$ is the electron charge. We note that on the one hand, charged and neutral grains couple to the electromagnetic field via collisions with the electrons and ions and on the other hand, charge fluctuation modifies this field \citep{V94, BP94, V17}. Thus, dust–-plasma coupling is responsible for some of the novel collective features in a dusty medium \citep{PV06, PV09, P99, D95}. The collision between the plasma, neutrals and dust grains not only causes the dissipation of the high frequency waves but can also help the dissipationless propagation of the low frequency fluctuations. For example, if various collision frequencies are higher than the dynamical frequency of interest then collision will move the bulk medium (which is a sum of the plasma, dust and neutral particles) together. In such a scenario collision facilitates undamped propagation of the wave \citep{P07a, PV07b, PVS12, PVI13}. However, in the opposite limit, when the collision frequencies are smaller than the dynamical frequency of interest, collision causes damping of the waves.

The $80-–120\, \mbox{km}$ region of Earth{}\'s ionosphere is weakly ionised and weakly magnetized with the neutral number density ($> n_n = 10^{14}\,\mbox{cm}^{-3}$ at $80\,\mbox{km}$) far exceeding the ion number density ($n_i = 10^{3}\,\mbox{cm}^{-3}$). The sub--visible small dust ($\sim \mbox{nm}$) number density could be similar to the ion number density. Adding to this complexity is the role of the ambient magnetic field. The dynamical processes in the ionosphere are strongly controlled by the coupling of the largely neutral D and lower E region to the magnetic field. This coupling is facilitated by the frequent collisions between the neutrals and the dusty plasma particles which transmits the Lorentz force to the neutrals. It is pertinent to recall here that in the past the D—-layer was considered either a purely neutral layer \citep{BT96} or, the role of dust in the mesospheric plasma dynamics was completely ignored \citep{AF07}. But as is clear from the above description, the abundance of tiny charged grains may well exceed the electron abundance in the D-—region and thus the plasma transport properties which neglects the presence of grains at this altitude is incomplete.  

The role of dust in modifying the plasma conductivity is well known in the space \citep{S11, Y16} and astrophysical environment \citep{PV07b, WN99}. The presence of dust not only affects the ionization structure of the plasma but also the gas phase abundances \citep{WN99, W07}. Further, owing to the large mass and size distribution, the grains can couple directly as well as indirectly to the magnetic field. Thus the charged grains not only modifies the ambipolar time–-scale, but may also give rise to the Hall diffusion \citep{WN99}. By ambipolar diffusion here we imply the diffusion of the magnetic field against the sea of neutrals due to the relative slippage of the frozen--in ions against the neutrals \citep{S78}. This is different from the ambipolar diffusion of the plasma particles against the electric field \citep{H78}. As we shall see, the mesosphere can be described in the framework of magnetohydrodynamics (with Ohm, ambipolar and Hall diffusion operating at various scale heights). This approach is different from the usual electrodynamics approach in which the magnetic field is assumed static. As has been noted by Parker (2007), in the magnetohydrodynamic approach, the bulk velocity of the plasma fluid and the magnetic field is the primary variable whereas in the electrodynamic approach, electric field and current is the primary variable. Both paradigm may sometimes arrive at the same conclusion \citep{L14}. In the present work, we shall adopt the magnetohydrodynamic approach and first explore relative importance of the various diffusivities before investigating the wave propagation in the dusty mesospheric layer.    

We shall provide an expression for the generalized Ohm{}\'s law in the next section where the relative importance of Ohm, Hall and ambipolar diffusion is also discussed . In section III we describe the effect of Hall and Ohm diffusion on the propagation of low frequency waves. It is shown that the long wavelength waves may suffer significant damping if most of the electrons have been moped by the dust. In section IV discussion of the result along with a brief summary is presented. 

\section{Formulation}
 We shall assume a weakly ionized medium consisting of electrons, ions, charged grains and neutral particles. In order to better elucidate the role of the charged dust, we shall assume that the grains have same size and ignore their size distribution. Likewise, for simplicity the difference between the molecular and cluster ions will be neglected. Since the D-—region is weakly ionized the inertia and the thermal pressure terms in the plasma momentum equations can be neglected.  

The most accurate way of calculating the transport properties of a gas is to employ Chapman—-Enskog method which was applied to the ionosphere by Cowling \citep{C45}. This method is analytically involved and thus we shall adopt a much simpler {\it free--path} method similar to the one employed by Baker and Martyn \citep{BM53} for the ionosphere. Free—-path theory assumes that the gas is not accelerating, i.e. it makes no distinction between the values of the mean velocity at the beginning of the free path and at the instant considered. Thus the charged particles drift through the sea of neutrals due to instant Lorentz force, i.e. 
\bq 
0 = - q_j\,n_j\,\left(\E' + \frac{\vj\cross \B}{c}\right) -
\rho_j\,\nu_{jn}\,\vj\,. 
\label{eq:em1} 
\eq
Here $ q_j\,n_j\,\left(\E' + \vj\cross \B/c\right)$ is the Lorentz force and $\E' = \E + \vn\cross \B / c$ is the electric field in the neutral frame with $\E$ and $\B$ as the electric and magnetic fields respectively,  $n_j\,,\v_j$ is the number density and velocity,  $q_j $ is the charge on the plasma particles and dust and $c$ is the speed of light.  The last term on the right hand in the above equation is the collision momentum exchange term in the neutral frame of reference. The electron–-neutral and ion—-neutral collision frequencies are \citep{K89}
\begin{eqnarray}
\nu_{en} &=& 5.4\times 10^{-10}\,n_{n}\,T_e^{1/2}\,,\nonumber\\
\nu_{in} &=& 2.6\times 10^{-9}\,n_{n}\,A^{-1/2}\,.
\end{eqnarray}
Here $A$ is the mean neutral molecular mass in atomic mass units. Assuming $A=30$ above collision frequency can be written in the following form \citep{PV11}
\begin{eqnarray}
\nu_{en} &=& 7.64\times 10^{4}\,n_{n+13}\,T_{+200}^{1/2}\,,\nonumber\\
\nu_{in} &=& 4.7\times 10^{3}\,n_{n+13}\,A_{30}^{-1/2}\,.
\end{eqnarray}
Here $A_{30} = A/30$, $n_{n+13} = n_n/10^{13}\,\mbox{cm}^{-3}$ and $T_{+200} = T/200 K$. Since the dust-neutral collision rate is \citep{NU86}
\bq
<\sigma\,v>_{dn} = 7.1 \cross 10^{-9}\, T_{+200}^{\frac{1}{2}}\,a_{-7}^{2}\quad
\mbox{cm}^3\,\mbox{s}^{-1}\,,
\label{eq:cf1}
\eq
with $a_{-7}=a/10^{-7} \mbox{cm}$, the dust-—neutral collision frequency
becomes
\bq
\nu_{dn} =\left(\frac{m_n}{m_d}\right)\,n_n\,<\sigma\,v>_{dn} \simeq 237\,n_{n+13}\,.\,,
\eq
Here we have assumed $m_n = 20\,m_p$ with $m_p= 1.67\times10^{-24}\,\mbox{g}$ and $m_d = 10^{-20}\,\mbox{g}$ for the dust mass density $\rho_d = 2\,\mbox{gm} / \mbox{cm}^{-3}$.

We shall define the plasma Hall parameter
\bq
\beta_j = 
\left(\frac{\omega_{cj}}{\nu_{jn}}\right)\,,
\eq
as the ratio of the cyclotron $\omega_{cj} = q_j\,B/m_j\,c$ to the collision $\nu_{jn}$ frequencies. The value of this parameter is a measure of how well the magnetic field is coupled to the neutral matter.

\begin{figure}
\includegraphics[scale=0.30]{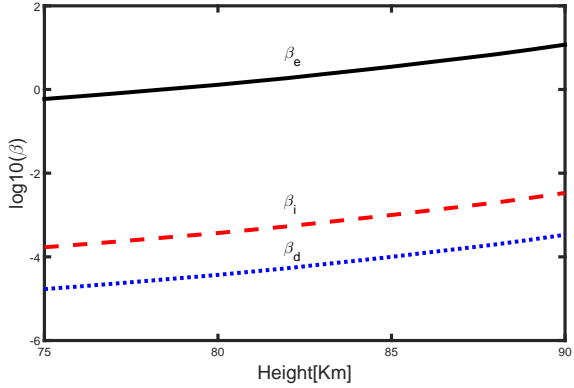}
\caption{The variation of Hall $\beta$ which is a ratio of the cyclotron to the plasma—-neutral collision frequencies is shown against the altitude in the above figure. The neutral number density for the summer mesosphere has been taken from \citep{L99}.}
\label{fig:FM1}
\end{figure}
Since $\omega_{ce} = 5.3\times10^6\,B_{0.3}\,\persec$, $\omega_{ci} = 10^2\,B_{0.3}\,\persec$ and $\omega_{cd} = 0.5\,B_{0.3}\,\persec$ for a nm--sized grain with $Z = -1$,  various Hall beta become
\begin{eqnarray}
\beta_e &\simeq& 70\,T_{+200}^{-\frac{1}{2}}\,n_{n+13}^{-1}\,B_{0.3}\,,\nonumber\\
\beta_i &\simeq& 2\times10^{-2}\,A_{30}^{\frac{1}{2}}\,n_{n+13}^{-1}\,B_{0.3}\,,\nonumber\\
\beta_d &\simeq& 2\times 10^{-3}\, a_{-7}^{-2}\,T_{+2}^{-\frac{1}{2}}\,n_{n+13}^{-1}\,B_{0.3}\,,
\label{eq:hbet}
\end{eqnarray} 
where $B_{0.3} = B/0.3\,\mbox{G}$. Clearly depending on the value of Hall $\beta$ the charged grains may couple directly or indirectly to the ambient magnetic field. In Fig.~(\ref{fig:FM1}) the variation of Hall $\beta$, Eq.~(\ref{eq:hbet}) is plotted against the altitude. We see from the figure that the electrons are magnetized above $78\,\mbox{km}$ altitude whereas ions and nanometre sized or, larger grains are
uncoupled to the magnetic field (which we shall also call unmagnetized for short) in the entire mesosphere. Therefore, the ions and dust grains can only indirectly couple to the magnetic field since in this case $\beta_d \ll \beta_i<1$.
 
The current $\J$ and the electric field $\E$ are related via conductivity tensor $\tenq{\sigma}$
\bq
\J = \sigma_{\parallel}\, \E'_{\parallel}+ \sigma_{P}\, \E'_{\perp}' + \sigma_H\,\E'\cross \hB\,,
\label{eq:Olw}
\eq
where $\hB = \B/|B|$ and the parallel, Pedersen and Hall conductivities can be written as \citep{W07}
\begin{eqnarray}
\sigma_{\parallel} &=& \left(\frac{e\,c}{B}\right)\,\sum_j n_j\,|Z_j|\,\beta_j\,,\nonumber\\
\sigma_P &=& \left(\frac{e\,c}{B}\right)\, \sum_j\frac{ n_j\,|Z_j|\,\beta_j}{1+\beta_j^2}\,,\nonumber\\
\sigma_H &=& - \left(\frac{e\,c}{B}\right)\, \sum_j\frac{ n_j\,Z_j}{1+\beta_j^2}\,,
\label{eq:cond1}
\end{eqnarray} 
where $Z_j = \pm 1$ is the sign of the charged particle. The Ohm ($\eta_O$), Hall($\eta_H$) and ambipolar ($\eta_A$) diffusivities are \citep{W07}
\begin{eqnarray}
\eta_O = \left(\frac{c^2}{4\,\pi}\right)\frac{1}{\sigma_{\parallel}}\,\,,\,
\eta_{A} = \frac{c^2}{4\,\pi}\left( 
\frac{\sigma_P}{\sigma_{\perp}^2} - \frac{1}{\sigma_{\parallel}}
\right)\,,\,
\nonumber\\
\eta_H = - \left(\frac{c^2}{4\,\pi}\right)\frac{\,\sigma_H}{\sigma_{\perp}^2}\,,
\label{eq:diff} 
\end{eqnarray}
where $\sigma_{\perp} = \sqrt{\sigma_{P}^2 + \sigma_{H}^2}$ and the Pedersen diffusivity is
\bq
\eta_P = \eta_O + \eta_A\,.
\eq   
Conductivity $\sigma$ has the unit $1/s$ in Eq.~(\ref{eq:cond1}). Similarly, the diffusivity has the unit $\mbox{cm}^2/\mbox{s}$ in Eq.~(\ref{eq:diff}).
\begin{figure}
\includegraphics[scale=0.30]{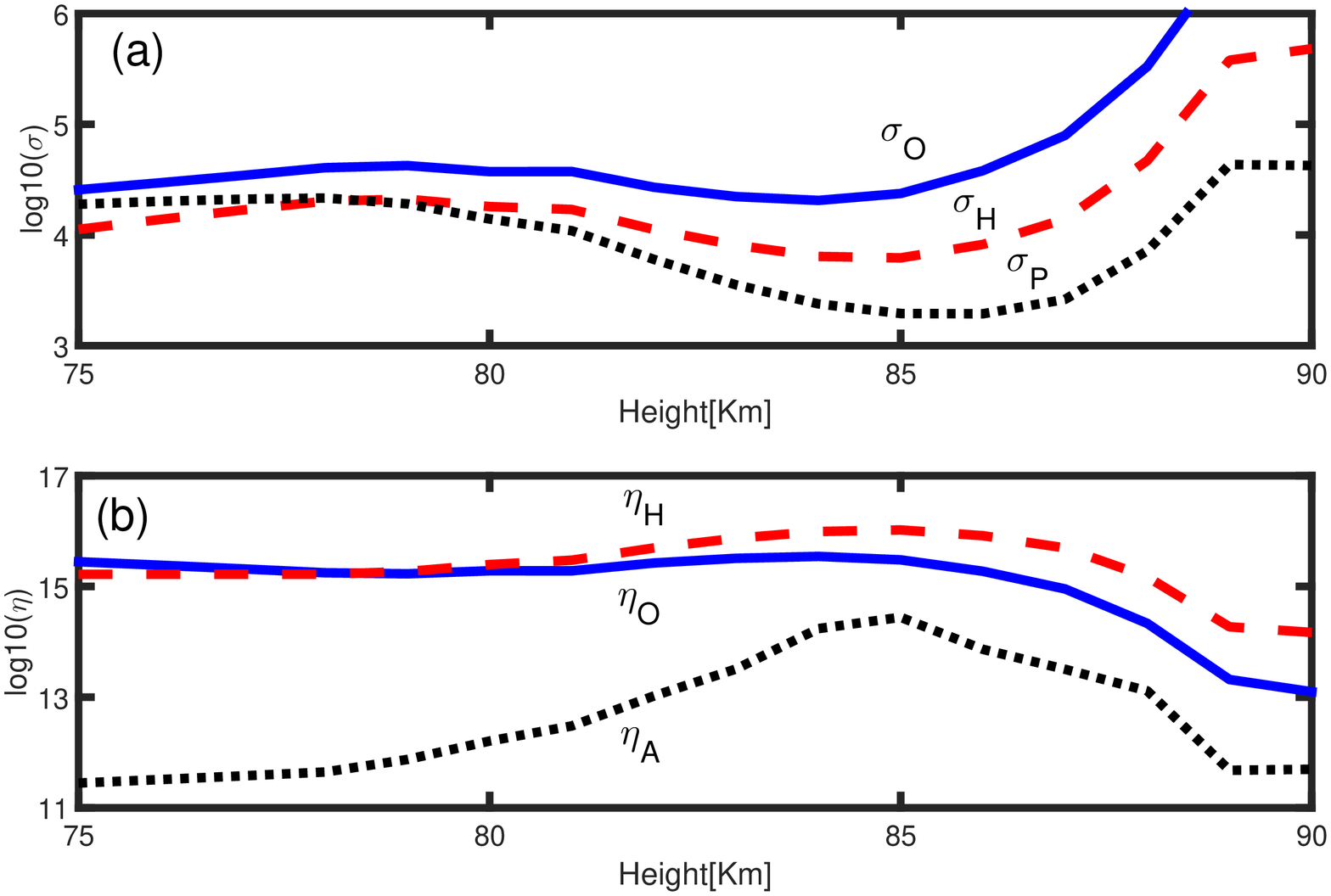}
\caption{The plasma conductivities ($1/s$) [panel (a)] and corresponding diffusivities ($\mbox{cm}^2/\mbox{s}$) [bottom panel (b)] is plotted against the PMSE altitude where the electron bite out occurs. Data is taken from ECOMA/MASS campaign [Fig.~5 \citep{R09}]. The subscript $H\,,O\,,A$ and $P$ in the figure stands for Hall, Ohm, ambipolar and Pedersen.}
\label{fig:FM2}
\end{figure}
In Fig.~\ref{fig:FM2}(a) we plot various conductivities closely corresponding to the PMSE altitude ($78-—80\,\mbox{km}$) where electron bite out occurs. The data from ECOMA/MASS campaign [Fig.~(5)] of \citep{R09} have been used for the above plot. The corresponding diffusivities in Fig.~[\ref{fig:FM2}(b)] suggest that the Hall and Ohm diffusion are comparable below $80\,\mbox{km}$ and Hall dominate beyond this altitude. The ambipolar diffusion is several orders of magnitude smaller than both the Ohm and Hall diffusion. Thus the Pedersen diffusivity at this altitude is solely due to Ohm diffusion. Clearly, Hall is the dominant diffusion in the electron bite—-out region. Although we have used ECOMA/MASS campaign data the dominance of Hall over other diffusion is a generic feature of the upper D—-region. 

How important is the presence of grains in the D-—region for the Hall diffusion? We know that the electron and ion densities are nearly equal after about $\sim 90\,\mbox{km}$ and the dust has very little role in affecting the plasma conduction properties above this altitude. However, below this altitude, the dust grains play an important role in the magnetic diffusion. In order to delineate the role of dust grains, we rewrite the Hall and Pedersen conductivities [Eq.~(\ref{eq:cond1})]
as
\begin{eqnarray}
\frac{\sigma_H}{\left(\frac{e\,c\,n_e}{B}\right)} &=&  \frac{1}{1+\beta_e^2} - \left(\frac{n_i}{n_e}\right)\, \frac{1}{1+\beta_i^2} + \left(\frac{n_g}{n_e}\right)\, \frac{1}{1+\beta_g^2}\,,\nonumber\\
\frac{\sigma_P}{\left(\frac{e\,c\,n_e}{B}\right)} &=&  \frac{\beta_e}{1+\beta_e^2} + \left(\frac{n_i}{n_e}\right)\, \frac{\beta_i}{1+\beta_i^2} + \left(\frac{n_g}{n_e}\right)\, \frac{\beta_g }{1+\beta_g^2}\,.
\nonumber\\
{}
\label{eq:CEx}
\end{eqnarray}

In the lower D-—layer ($\ll 75\,\mbox{km}$), the collision frequencies are so high (in comparison with the respective cyclotron frequencies) that the electrons, ions and dust grains are unmagnetized, i.e. $\beta_g \ll \beta_i \ll \beta_e\ll 1$. In the absence of dust, the conductivity is isotropic since $\sigma_{\parallel} \approx \sigma_{P}$ and $\sigma_{H} \approx 0$. However, in the presence of grains this isotropy is lost since
\begin{eqnarray}
\sigma_{\parallel} \approx \sigma_{P} &\approx& \left(\frac{e\,c}{B}\right) n_e\,\beta_e\,,\nonumber\\
\sigma_H  &\approx&\left(\frac{e\,c\,n_i}{B}\right)\,\beta_i^2\Big[1-\left(\frac{n_g}{n_i}\right)\,\left(\frac{\beta_g}{\beta_i}\right)^2\Big]\,,  
\label{eq:sigD}
\end{eqnarray}
where in the above expression for the Hall conductivity $\sigma_H$ we have assumed $\beta_g\ll\beta_i<1<\beta_e$ and
\bq
\left(\frac{n_i}{n_e}\right) > \frac{1}{\beta_i^2}\,.
\label{eq:qnc}
\eq
Making use of (\ref{eq:hbet}), Eq.~(\ref{eq:qnc}) gives $n_i/n_e>2.5\times\left(10^3-—10^4\right)\,$ at $80—-90\,\mbox{kms}$ altitude. This inequality may be easily satisfied in the electron bite out region where most of the electrons are mopped by the dust grain.

\begin{figure}
\includegraphics[scale=0.30]{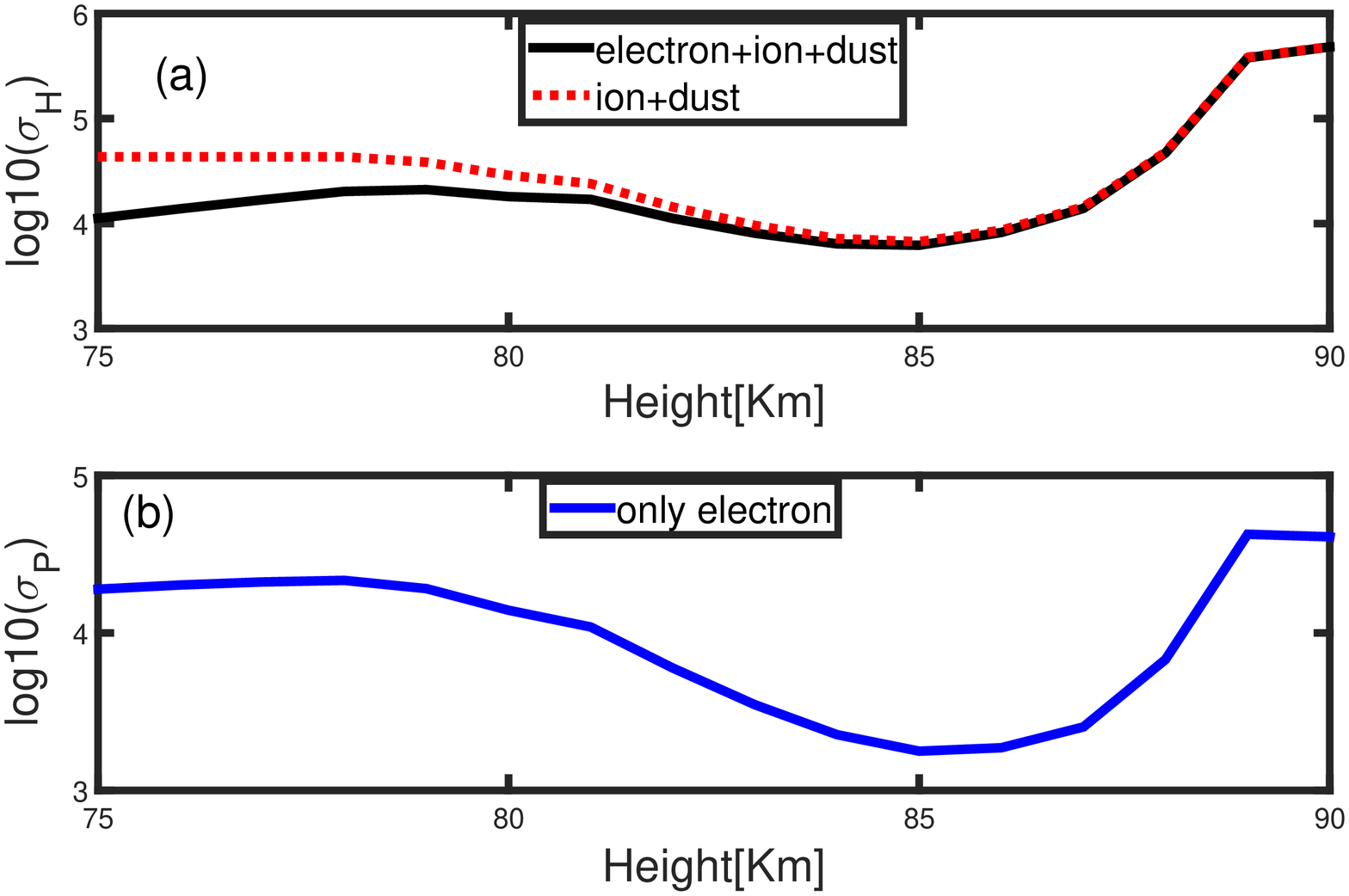}
\caption{The Hall conductivity ($1/s$) at PMSE altitude with and without the electron term in Eq.~(\ref{eq:CEx}) is plotted in the top panel (a). The Pedersen conductivity due only to the electron term is plotted in panel (b).}
\label{fig:FM3}
\end{figure}
In Fig.~(\ref{fig:FM3}) we plot the Hall conductivity in the top panel with and without the electron term in Eq.~(\ref{eq:CEx}). It is clear from the figure that beyond $78-80\,\mbox{km}$ altitude in the PMSE region, the Hall conductivity is determined by the relative transverse drift between the weakly magnetized  ions ($\beta_i < 1$) and unmagnetized dust ($\beta_g\ll1$).  Below $\sim 78-80\,\mbox{km}$, the Hall conductivity is due to the relative drift between the unmagnetized  ions and weakly magnetized electrons. Note that at this altitude since $\beta_e<1$ the electron—-neutral collision easily dissipates this relative drift (current). Thus both the Hall and Ohm diffusion are comparable below $\sim 80\,\mbox{km}$ in Fig.~(\ref{fig:FM2}). 

Clearly the origin of the Hall effect in the lower and upper D-—region is quite different. Whereas it is the relative transverse electron-—ion drift in the lower D-—region, it is the ion—-charged dust drift in the upper D-—region that leads to the Hall conductivity. When the electron is also strongly unmagnetized in the lower D-—region [$\beta_e\ll1$, not shown in Fig.~(\ref{fig:FM1})], the Hall conductivity disappears. The Pedersen conductivity [bottom panel Fig.~(\ref{fig:FM3})] is determined mainly  by the electron term in Eq.~(\ref{eq:CEx}) since there is very little difference with the $\sigma_P$ curve in the top panel of Fig.~(\ref{fig:FM2}). 

Inverting generalized Ohm's law Eq.~(\ref{eq:Olw}) in terms of electric field $\E$ gives
\bq
\frac{c^2}{4\,\pi}\E^{\prime} = \eta_O\,\J_{\parallel} + \eta_H\,\J\cross\b + \eta_P\,\J_{\perp}\,.
\label{eq:ef}
\eq
Taking curl of Eq.~(\ref{eq:ef}) and using of Maxwell{}\'s equation $c\,\curl\E = -\partial_t \B$, we get following induction equation 
\begin{eqnarray}
\delt \B = \curl\,\left[\left(\v\cross\B\right) - \frac{4\,\pi\,\eta_O}{c}\,\J + \frac{4\,\pi\eta_A}{c}\, \left(\J\cross\hB\right)\cross\hB  \right. \nonumber\\
\left. - \frac{4\,\pi\,\eta_H}{c}\,\left(\J\cross\hB\right)\right]\,. 
\label{eq:ind} 
\end{eqnarray}
Here $\v$ is the bulk fluid velocity \citep{PV11}. 

As the size distribution of the dust in the mesosphere is skewed towards nanometre (nm) or, smaller particles, it is quite plausible that the fine, nm sized dust may mop up all the electrons in the plasma leaving behind the larger dust ($\sim$ few $10$s of nanometre) to collect the more sluggish positive ions. This may happen especially when the fraction ionization (ratio of the electron and neutral number densities) in the partially ionized gas is very low, i.e. when the plasma is weakly ionized. The observation of the nighttime polar mesosphere over Kiruna, Sweden confirms such an expectation \citep{R05}. Thus when the fine dust particles ($< 50\,\mbox{amu}$) mops up most of the electrons and the positive charge resides on the micron and sub—-micron sized grains (See Fig.~5 of \citep{R05}), it is the relative transverse drift (with respect to the magnetic field) of the negatively charged and magnetized fine dust powder against the {\it almost stationary} unmagnetized or, weakly magnetized positive dust that will give rise to the Hall effect. Since the direction of the Hall electric field in this case will be opposite to the case discussed above where it was the plasma particles that was drifting against the negatively charged dust, the sign of the Hall term in the induction equation (\ref{eq:ind}) will become positive. Such a reversal of sign of the Hall term is well known in the dusty plasmas \citep{PVI13} and has recently been investigated in the plume region of the Saturn{}\'s moon Enceladus \citep{YL18}.

It is pertinent to ask at this stage, what is the spatial scale over which the magnetic field couples to the plasma in the D-—region. In order to answer this question we need to compare the advective and diffusive terms in the induction equation. Thus from Eq.~(\ref{eq:ind}) we get $L \sim \eta_H/v$. Assuming $v \sim v_A \sim 0.1-1 \mbox{km}/\mbox{s}$ for $\eta_H \sim 10^{14}\,\mbox{cm}^2/\mbox{s}$ [Fig.~(\ref{fig:FM2})] the characteristic length scale $L$ turns out to be about $\sim 10^3-10^4\,\mbox{km}$. Thus we conclude that the Hall diffusion of the magnetic field is important only for the ultra—-low frequency waves corresponding to the planetary waves of similar wavelength \citep{A05}. For the shorter wavelengths, magnetic diffusion in Eq.~(\ref{eq:ind}) should be safely neglected. However, the ideal MHD framework of the D—-region plasma (which is partially ionized) is not the same as the ideal MHD of fully ionized plasmas \citep{P16}.
 
As can be seen from Eqs.~(\ref{eq:diff}) and (\ref{eq:sigD}), the ratio of the ambipolar and Hall diffusion coefficient is 
\bq
\left|\frac{\eta_A}{\eta_H}\right| \sim \left|\left( 
\frac{\sigma_P}{\sigma_{\perp}^2} - \frac{1}{\sigma_{\parallel}}
\right) \frac{\sigma_{\perp}^2}{\sigma_H}\right| \approx 0\,,
\eq
and thus the ambipolar diffusion in the induction Eq.~(\ref{eq:ind}) is unimportant. We shall note yet again that here we are talking about the ambipolar diffusion of the magnetic field against the sea of neutrals. This is different than the ambipolar diffusion of the electrons against the electric field. The presence of dust particles may significantly reduce the electron density, resulting the slowing down of electron diffusion which may facilitate the survival of plasma structures over long ($10$s of minutes to hours) lifetime \citep{RL03}. However, like the usual MHD framework, the present, dusty magnetohydrodynamics framework does not dwell upon the electric diffusion. 

The ratio of the Hall and Ohm diffusion coefficient is
\bq
\left|\frac{\eta_H}{\eta_O}\right| \sim \left|\frac{\sigma_H}{\sigma_{\parallel}}\right|\sim \frac{\left(1+P\right)\,\beta_i^2}{\beta_e}\,\left|1-\left(\frac{n_g}{n_i}\right)\,\left(\frac{\beta_g}{\beta_i}\right)^2\right|\,,
\eq
where
\bq
P = \frac{Z\,n_g}{n_e}\,,
\eq 
is Havnes parameter. In the electron bit—-out regions where most of the electrons has been mopped by the dust, we see that $n_g \sim n_i$ (owing to the quasi-—neutrality) and thus, it is not guaranteed that the Hall will dominate Ohm diffusion. Clearly, both the Ohm and Hall diffusion may operate on the same footing in the lower mesosphere conforming what we already know from Fig.~(\ref{fig:FM2}). As we shall see in the next section whereas Hall will cause the low frequency, long wavelength left and right circularly polarized cyclotron and whistler waves the Ohm will cause the damping of such waves.
\section{Waves in the mesosphere}

Defining the  bulk fluid mass density and velocity as $ \rho \approx \rho_n\,, \v \approx \v_n$ and after assuming a barotropic relation $P = c_s^2\,\rho$ where $c_s$ is the sound speed we linearize the following equations  
\bq
\delt\rho + \grad\cdot\left(\rho\,\v\right) = 0\,, 
\label{eq:cont1} 
\eq 
\bq
\rho\,\frac{d\v}{dt} =  - \nabla\,P + \frac{\J\cross\B}{c}\,,
\label{eq:meq} 
\eq
together with the induction Eq.~(\ref{eq:ind}) around a homogenous background $\B_0,\,\rho_0,\,P_0$ and get the following equations
\begin{equation}
\omega\,\drho - \rho\,\k\cdot\dv = 0\,,\nonumber\\
\end{equation}
\begin{equation}
\omega\,\dv = c_s^2\left(\frac{\k\cdot\dv}{\omega}\right)\,\k
+ \Big[\left(\b\cdot\dB\right)\,\k-\left(\k\cdot\b\right)\dB \Big]\va^2\,,
\label{ml1}
\end{equation}
where $\va = B/\sqrt{4\,\pi\,\rho}$ is the \alf speed in the bulk fluid and $\dB=\delta\B/|B|$. We define $\bar{\omega}^2 = \omega^2 - k^2\,c_s^2$, dot equation
(\ref{ml1}) with $\k$, and use $\k\cdot\dB = 0$ to write
\begin{equation}
\k\cdot\dv = \left(\frac{\omega_A}{\bar{\omega}}\right)^2\,\omega\, \left(\b\cdot\dB\right).
\label{cnt}
\end{equation}
Here $\omega_A = k\,\va$ is the \alf frequency. Making use of Eq.~(\ref{cnt}) Eq.~(\ref{ml1}) can be written as
\bq\omega\,\dv = \left[
\left(\frac{\omega}{\bar{\omega}}\right)^2\left(\b\cdot\dB\right)\,\k 
-\left(\k\cdot\b\right)\dB\right]\,\va^2\,.
\eq
The linearized induction equation after setting $\eta_A=0$ can be written as
\begin{eqnarray}
\left(\omega-i\,k^2\,\eta_O\right)\,\dB =\left[\left(\k\cdot\dv\right)\hB 
- \left(\k\cdot\hB\right)\dv \right] 
\nonumber\\
-  i\,\eta_H\,\left(\k\cdot\hB\right) \k\cross\dB\,.
\label{eq:indx1} 
\end{eqnarray}
Defining $\kh\cdot\hB = \cs\,,$($\kh=\k/|\k|$) and diffusivities $D_O = k^2\,\eta_O$, $D_H = k^2\,\eta_H$ and eliminating $\dv$ and $\k\cdot\dv$ from the above equation  yields 
\begin{eqnarray}
\left[\omega^2 –- i\,D_O\,\omega - \omega_A^{2}\,\cos^2\theta\right]\dB = \left(\frac{\omega}{\bar{\omega}}\right)^2\,\omega_A^2\,
\nonumber\\
\times\left(\dB\cdot\hB\right)\,\left(\hB - \kh\,\cs \right)
- i\,D_H\,\omega\,\cs\,\left(\hat{\k}\cross\dB\right)\,.
\label{ME}
\end{eqnarray}
After some straightforward algebra we get following dispersion relation 
\begin{eqnarray}
\left[\omega^2 \left( 1 - \frac{\omega_A^2}{\bar{\omega}^2}\,\sin^2\theta\right)- i\,D_O\,\omega - \omega_A^{2}\,\cos^2\theta  \right]
\nonumber\\
\times
 \left(\omega^2 –- i\,D_O\,\omega - \omega_A^{2}\,\cos^2\theta\right)
- D_H^2\,\omega^2\,\cos^2\theta = 0\,.
\label{eq:DER}
\end{eqnarray}
This dispersion relation acquires a familiar form \citep{PV09, PVI13} when wave is propagating along the ambient magnetic field ($\theta = 0$)
\begin{equation}
\omega^2 –- i\,k^2\,\eta_O\,\omega
- \omega_A^2 = \pm \eta_H\,k^2\,\omega\,.
\label{eq:WN}
\end{equation}
In the Ohm limit the solution is
\bq
\omega = \frac{i\,k^2\,\eta_0}{2} \pm \omega_A\,\left(1-\frac{k^2\,\eta_O^2}{4\,\va^2}\right)^{1/2}\,,
\eq
which for small magnetic diffusion describes an \alf wave propagating at frequency $\omega \approx \omega_A$, and which is experiencing damping at a rate $k^2\,\eta_O/2$. The short wavelength fluctuations when the magnetic field diffusion speed  is of the order or, larger than twice the \alf speed, i.e.   $k\,\eta_O > 2\,\va$ do not survive as there is no oscillatory solution, only damping.  When $\theta = \pi /2$, the Hall term drops out from Eq.~(\ref{eq:DER}). The damped magnetosonic mode is the solution of the dispersion relation.

Eq.~(\ref{eq:DER}) in the Hall limit becomes 
\begin{eqnarray}
\left(\omega^2 - \omega_A^2\,\cos^2\theta\right)^2 - \left(\frac{\omega^2 - \omega_A^2\,\cos^2\theta}{\omega^2 -– k^2\,c_s^2}\right)\,\omega^2\,\omega_A^2\,\sin^2\theta 
\nonumber\\
=  \left(k^2\,\eta_H\right)^2\,\omega^2\,\cos^2\theta\,.
\label{eq:d1f}
\end{eqnarray}
For the waves propagating transverse to the ambient field, i.e. $\theta = \pi / 2$, the dispersion relation (\ref{eq:d1f}) gives
 \bq
\omega^2 = k^2\,\left(c_s^2 + \va^2\right)\,,
\label{eq:ms1}
\eq 
which is the usual magnetosonic branch. For the waves propagating along the background magnetic field ($\theta = 0$) above dispersion relation (\ref{eq:d1f}) gives
\bq
\omega = \left(\frac{\omega_W}{2}\right)\left(
1 \pm \sqrt{1\pm4\,\frac{\omega_A^2}{\omega_W^2}}\right)\,.
\label{eq:whist}
\eq
where
\bq
\omega_W = k^2\,\eta_H\,, 
\eq
is the whistler frequency.  The positive and negative sign inside the square bracket in (\ref{eq:whist}) corresponds to the left and circularly polarized waves owing to the handedness of the Hall effect.

Note that the Eq.~(\ref{eq:whist}) is the nonlinear dispersion relation for the exactly parallel (say along axis $z$), circularly polarized \alf waves, $\b = \exp{i\,\left(\omega\,\tau-k\,\xi\right)}$ \citep{P07a, PV07b}. Here $\tau$ and $\xi$ are stretched variables and $b = B_x+i\,B_y$. The weakly nonlinear and weakly dispersive waves propagating exactly along or, slightly oblique to the magnetic field satisfies complex derivative nonlinear Schrodinger (DNLS) equation. For exactly parallel case, the DNLS equation admits localized envelop soliton \citep{PVS08}
\bq
|b|^2 = 2\,\left(\frac{V}{\va}\right)\,\Big[\sqrt{2}\cosh\left(\frac{\xi-V\,\tau}{L}\right)-1\Big]^{-1}\,,
\eq  
where $V$ represents the velocity in the co—moving frame, $L=\eta_H/2\,V$ is the width and $2\,V/\va$ is the maximum amplitude of the soliton. Here we have assumed right circularly polarized waves.  For typical \alf speed $\sim 0.1-1\,\mbox{km}$ and $\eta_H \sim 10^{14}-10^{15}\,\mbox{cm}^2/\mbox{s}$ [Fig.~\ref{fig:FM2}(b)] the typical width of the soliton turns out to be $\sim 10^{3}-10^{5}\,\mbox{km}$. However,  the presence of Ohmic dissipation in the mesosphere may give shock—-like structure due to dissipation of the wave energy. 
\section{Discussion and summary} 
In a partially ionized medium in the presence of magnetic field the plasma conductivity becomes anisotropic. The three orthogonal components of this anisotropic tensor are (i) ambipolar, when the magnetic field is frozen in the electron fluid and slips through the neutrals, (ii) Hall-- when the plasma particles are partially coupled or, uncoupled to the magnetic field and, (iii) Ohm, when the plasma particles cannot drift with the neutrals owing to frequent collisions. In the mesosphere however since only the electron fluid is frozen in the magnetic field above $80\,\mbox{km}$ and the ions are either unmagnetized (in which case they move with the neutrals owing to frequent collisions) or, weakly magnetized (i.e. the relative slip between the ions and the neutrals is small) the ambipolar diffusion is negligible. Thus only the Hall and Ohm is important in the mesosphere. However, the origin of Hall effect is quite different below and above $80\,\mbox{km}$. Whereas below $80\,\mbox{km}$ altitude the Hall effect is due to the relative transverse drift between the weakly magnetized electrons and unmagnetized ions (the dust provides a stationary neutralizing background), above this altitude, the Hall effect is due to the relative drift between the weakly magnetized ions and unmagnetized dust. Note that the distribution of the dust in the mesospheric layer determines the sign of the Hall effect. If the negative charge largely resides on the fine, nanometre sized dust grains and positive charge resides on the more massive (few $10$s of nanometre) grains \citep{R05}, then the sign of the Hall electric field will be opposite to the previous case as in this case the transverse drift of the magnetized negative fine dust against the more massive unmagnetized positive dust causes the Hall effect. The sign change of the Hall term in the induction equation (\ref{eq:ind}) may affect the polarization of the waves in the mesospheric plasma layers. 

As the Hall and Ohm diffusion are comparable below $80\,\mbox{km}$, the low frequency ($\omega \sim 10^{-4}-—10^{-5}\,\mbox{s}^{-1}$) fluctuations will be damped with the damping rate $k^2\,\eta_O \sim$ few minutes. Therefore, the ultra--low frequency waves will be heavily damped below $80\,\mbox{km}$ in the mesosphere. However, above $80\,\mbox{km}$ the mesosphere can host ultra--low frequency waves as Hall dominates Ohm. As a result the long wavelength, low frequency circularly polarized whistlers can propagate across the ionosphere with little or, no damping at all. Since these low frequency waves have very large wavelength they may couple to the magnetosphere. For example, the fundamental frequencies at which magnetosphere resonates is 1.3 and 1.9\,mHz \citep{S91}. Equating this frequency with the whistler $k^2\,\eta_H$  gives  $\lambda \sim 10^3-—10^{4}\,\mbox{km}$. Therefore, it is quite likely that the generation of the low frequency waves in the lower E—-region or, upper mesosphere is responsible for the mesospheric resonant tuning.  

The present magnetohydrodynamic formulation of the weakly ionized plasma 
is complementary to the usual electrodynamic formulation. Whereas in the present magnetohydrodynamics framework the bulk plasma flow velocity ($\v$) and magnetic field ($\B$) are the primary variable, it is the current ($\J$) and electric field ($\E$) that are often used to describe the ionosphere dynamics. Therefore, the present framework does not deal with the diffusion of the electron against the electric field, which also is called the ambipolar diffusion. We have seen that for the mesospheric parameters, ambipolar diffusion of the magnetic field is unimportant. The effect of dust on the electron diffusion in the presence of {\it external} electric field cannot be addressed in this framework.  Importantly since the electric field is assumed to have its origin in the {\it external} events (e.g. solar wind \citep{L90} and references therein), it cannot penetrate the quasi--neutral plasma except in a very thin boundary layer of the order of the Debye Length. Therefore, the limited penetration of such an electric field in the plasma will not be able to set the convective motion and drive currents \citep{V12}.

Following is the summary of the results.

1. The Ohm and Hall is the main components of the plasma conductivity in the mesosphere.\\  
2. Whereas Ohm and Hall are comparable in the lower mesosphere, Hall dominates Ohm above $80\,\mbox{km}$.\\
3. The physical mechanism of Hall effect differs in the lower and upper mesosphere. The relative transverse (with respect to the ambient magnetic field) drift between the weakly magnetized electrons and unmagnetized ions causes the Hall in the lower mesosphere. In the upper mesosphere the relative drift between the weakly magnetized ions and unmagnetized nano metre or, larger dust causes Hall effect.\\
4. If the negative change in the mesosphere is carried, mainly by the fine dust particles and the positive charge is carried by the more massive almost immobile dust, the Hall effect in this case can be caused by the relative transverse drift of the magnetized fine dust against the positively charged dust. The sign of the Hall effect in this case will be opposite of the previous case. Thus the sign of the Hall effect not only depends on the direction of the of the local magnetic field but also on the nature of the charge carrier in the plasma.\\   
5. The mesosphere can host low frequency long wavelength waves that may propagate undamped through the ionosphere. 

\vspace{0.2in}
{\bf{ACKNOWLEDGEMENTS}}
The work of SVV was partially supported by the Government of Russian Federation (Grant 08-08). The support of the Australian Research Council is gratefully acknowledged for the present work.
\vspace{0.2in}

%
\end{document}